# A Matching Mechanism with Anticipatory Tolls for Congestion Pricing

J Ceasar Aguma(jaguma@uci.edu), Amelia C Regan(aregan@ics.uci.edu)
Department Of Computer Science, UC Irvine.

This paper presents a matching mechanism for assigning drivers to routes where the drivers pay a toll for the marginal delay they impose on other drivers. The simple matching mechanism is derived from the RANKING algorithm for online bipartite matching proposed by Karp et al. [8]. The toll, which is anticipatory in design, is an adaption of one proposed by Dong et al. [5]. Our research proves that the matching mechanism proposed here is pareto user-optimal and can be adapted to give network optimal results for the minimizing total social cost of travel.

## 1 INTRODUCTION

To some, taxing vehicle operators for road access seems unfair because it is clearly a regressive tax. However, it is highly logical because in addition to reducing road use, it could provide funds for improving roads and extending both the reach and convenience of transit services which should directly benefit less affluent travelers. A deeper look reveals that congestion pricing and related market mechanisms pose very complex economic problems. That is, establishing where in the networks to impose prices and what prices to pick to give road users incentives to limit their non-essential trips while also making essential road use available for middle and low income users.

Many economists have tried to put realistic numbers on the cost of congestion in terms of time lost in traffic and fuel wasted, but the true costs are much higher because of road injuries to drivers, cyclists and pedestrians and the short-term and long term enormous environmental impacts.

Despite the emergence of new Mobility-as-a-System (MaaS) services[9] in the past decade, both the number of private car owners and overall vehicle miles traveled (VMT) steadily increases every year. In fact, in most urban areas MaaS has increased VMT without decreasing car ownership because these services have drawn transit users out of those systems and increased the fraction of single occupancy vehicles around airports. Without new road pricing solutions we can confidently predict that road congestion will soon become the leading cause of air pollution and road accidents.

In this paper we propose the design of a market matching mechanism for as- signing drivers to travel routes in which space on congested routes is allocated to drivers willing to pay a fee which is equal to the cost of the delay imposed on other drivers. We show that our proposed matching mechanism is stable, strategy-proof and efficient.

### 1.1 Literature Review

Our paper was inspired by the recent work of Heller et al. [7]. That very recent paper presents a comprehensive literature review. Therefore, here we restrict ourselves to discussing the most related papers. We differ from some other researchers in that rather than explicitly consider drivers' individual value-of-time (VOT) we use willingness to pay tolls rather than selecting toll-free routes as a proxy for VOT. In this approach we differ from Arnott et al. [1], van den Berg and Verhoef [11] and many other researchers. Our work also draws on earlier work in Bernstein and El Sanhouri [3] which considers congestion pricing with an untolled alternative.

### 1.2 Summary of Contributions

We start off with a model for the route assignment problem under traffic congestion conditions in section 2, complete with a problem formulation and summary of the proposed algorithm. Where [7] uses an auction mechanism, we propose the employment of a matching mechanism with an anticipatory toll as an algorithmic solution in section 3. The simple matching mechanism proposed here is derived from the RANKING algorithm for online bipartite matching proposed by Karp et al.

[8]. The toll, which is anticipatory in design, is an adaption of one proposed by Dong et al. [5]. We use a neural network model to predict short term future traffic flow conditions and charge a toll that reflects those future conditions. The text goes on to prove that the matching mechanism proposed here is pareto user-optimal as defined by [12], strategy-proof as defined by Nisan et al. [10], and can be adapted to give network optimal results for minimizing total social cost of travel. In section 4, we compare this matching with the auction mechanism when both are applied towards traffic congestion pricing. A discussion follows in section 5.

## 2 MODEL

In this section, we present a problem definition, complete with theory of how the anticipatory toll is obtained, and a summary of the matching mechanism employed. As a prelude, the notation used is listed below.

Table 1. Table of Notation

| | | |
|---:|:---:|:---|
| $r$ | ≜ | Route $r$ |
| $d$ | ≜ | Driver $d$ |
| $q$ | ≜ | Timestep |
| $k_f$ | ≜ | Route threshold capacity for free flowing traffic |
| $k_t$ | ≜ | Route capacity at a time $t$ |
| $E_t$ | ≜ | Estimated Travel time at a specific time t |
| $E_f$ | ≜ | Estimated Travel time with free flowing traffic |
| $C_r$ | ≜ | Toll on route r |
| $C_d$ | ≜ | Toll paid by driver i |
| $X_t$ | ≜ | Traffic concentration/flow at time t |
| $A,B,F$ $\beta$ | ≜ | Congestion constants |
| $\alpha_d$ | ≜ | Driver d's willingness to pay toll(Equivalent to Value of Time) |
| $R(E,C)_r$ | ≜ | Route r's cost function |
| $U(E,C)_d$ | ≜ | Driver d's utility function |
| $P$ | ≜ | Penalty charge |

### 2.1 Problem Formulation

Consider a metropolis where $n$ drivers want to travel from a common origin $O$ to a common destination $D$, with $m$ routes available to the drivers. Note that a typical city will have many such O-D pairs, for example residential-work area pairs from which most travelers leave at common times.

Define a route r to have;

- Threshold Capacity, $k_f$, which is the route capacity that allows for free flowing traffic.
- A cost function, $R(E,C)_r = (E_t - E_f)C_r$

Additionally, Define a driver d to have;

- A Willingness-to-pay measure (Value of time), $\alpha_d$, a maximum toll the driver is willing to pay, with the assumption that all drivers prefer the fastest routes.
- A utility function, $U(E,C)_d = (E_f - E_t)C_d$, where;
  - $E_t = E_f[1 + A(\frac{k_t}{k_f})^B]$ As presented by Chen et al. [4].

$- C_r^t = C_r^{t-1} + \beta(X_{t+q} - X_t)$ As presented by Dong et al. [5].

Ideally, we would like to dynamically match drivers to routes in some way such that every driver $d$'s utility is maximized and while also minimizing total congestion cost for all routes.

Here, we propose a simple online matching algorithm;

---

**ALGORITHM 1:** Online Driver-Route Matching

---

**for** *Each driver that comes online* **do**

    (1) Rank all routes available to driver d by the projected utility $U(E, C)_d$

    (2) Assign top ranked route to driver $d$ with a deadline to accept the assignment

**end**

---

With a deadline to the assignment, drivers have an option to opt out of travel, however, if they do travel and use or switch to a route different from the assigned route, they get penalized with some charge; $P = C_r^t + F$, where $F$ is a fixed rate. Analysis in later sections will show that this simple matching mechanism is Pareto user-optimal and also minimizes the total congestion cost to within at least $1 - \frac{1}{e}$ of the network optimal. We will also show that prove that the matching is strategy-proof.

## 2.2 Anticipatory Toll

The cost of congestion on a route r is distributed among all travelling drivers in the form of an anticipatory toll that is calculated from future traffic conditions. Below, we provide a definition.

$$C_r^t = C_r^{t-1} + \beta(X_{t+q} - X_t)$$

The driver is then charged as follows;

$$C_d^t = \begin{cases} \frac{C_r^t}{k_t}; & if \ \ k_t - k_f > 0 \\ 0; & otherwise \end{cases} \tag{1}$$

Dong et al. [5] compare the effectiveness of a static toll, reactive toll, and an anticipatory toll. Their findings show that the anticipatory toll provides the best throughput.

*2.2.1 Calculating Future traffic flow ($X_{t+q}$).* We propose the use of convolution-long shortterm models (CNN-LSTM) to predict short term traffic flow. Asadi and Regan [2] provide and compare a few such model that takes an input a vector $X = X_0, X_1, \ldots, X_t$ of traffic flow conditions up until time $t$, then provides as output a traffic flow prediction at time $t + q$. We refer the reader to Asadi and Regan [2] for more detail.

## 3 MATCHING MECHANISM

Traffic congestion is a socioeconomic problem therefore any solution for congestion pricing has to be evaluated by how efficient it is from the perspective of road users and society as a whole. With that in mind, this section provides a proof that the matching algorithm proposed by this text is indeed pareto user-optimal and also achieves a fraction of the optimal total route network cost. We begin with an illustrative example of the algorithm.

## 3.1 Illustrative Example

Given a collection of $n$ drivers in the order in which they came online, $D = (d_1, d_2, \ldots, d_{n-1}, d_n)$ and $m$ available routes, $R = (r_1, r_2, \ldots, r_{m-1}, r_m)$ For each driver $d_i$;

(1) Rank all the routes such that, $U(E_{r_i}, C_{r_i})_{d_i} \geq U(E_{r_{i+1}}, C_{r_{i+1}})_{d_i}$, where $U(E_{r_i}, C_{r_i})_{d_i}$ is the utility driver $d_i$ would get from taking route $r_i$. Ties are broken by route capacities, $k_t$.

(2) Assign the top ranked route with a deadline within which the driver must begin their travel.

While it is intuitively clear that this user-optimal, a short proof is provided below for completeness.

## 3.2 Pareto Optimality

As a precursor to the proof, we provide a definition of Pareto Optimality.

Given $n$ users and $n$ resources, an allocation $X = (x_1, x_2, ........, x_{n-1}, x_n)$ is pareto optimal if any user $i$ can not obtain better utility in any other allocation $X' = (x'_1, x'_2, ........, x'_{n-1}, x'_n)$. That is, for each user $i$;

$$U_i(x_i) \geq U_i(x'_i)$$

with at least one user $j$ for whom $U_j(x_j) > U_j(x'_j)$.

*3.2.1 Proof.* Define the welfare of the matching $\mu$ produced by the algorithm for $n$ drivers to be the sum of all utilities,

$$W(\mu) = \sum_{i=1}^{n} U_i(\mu)$$

Let us assume there was a better matching $\mu'$ where at least one driver $d$ has a different assignment from that they get in $\mu$. This would imply that for $d$, $U_d(\mu') > U_d(\mu)$ which would be a contradiction to $\mu'$ being better because $W(\mu) > W(\mu')$

We can conclude that $\mu$ dominates all other matchings $\mu'$ and is therefore Pareto user-optimal.

## 3.3 StrategyProofness

A matching mechanism is strategy-proof if truth telling is a utility maximizing strategy, that is, the only way an agent can be guaranteed to get maximum utility is if they report true information.

For our matching, the driver is required to report the maximum toll they are willing to pay and get online when they are ready to travel. Lets consider two cases for the proof.

*3.3.1 Case 1: Driver gets online early to get better route.* In this scenario, we have a driver $d$ log on early so as to be higher in the driver queue. If a route is assigned and the driver does not start traveling before the deadline, the assignment will be terminated, in which case the driver gets $U_d = 0$. The only way they can get maximum possible utility is if they request a route when they are ready to travel within the deadline.

*3.3.2 Case 2: Driver under reports the maximum toll they are willing to pay .* Lets also consider a driver $d$ that reports their maximum toll to be $l$ but they are actually willing to pay a higher toll $h$. Driver $d$'s possible utility will be upper bounded by the possible utility with a toll of $a$, that is,

$$U(E, C)_d = (E_f - E_t)C_d \leq (E_f - E_t)l < (E_f - E_t)h$$

They would therefore get routes that do not maximize their utility. Further we make the assumption that reporting a higher maximum would be individually irrational because the driver would be assigned routes whose tolls they can not or will not pay. Therefore we do not explicitly include this as a separate case.

## 3.4 Network Optimality: matching for social good

This paper assumes that maximizing users' utility is a primary goal. However, the matching mechanism proposed here can easily be adjusted to control the negative social costs of traffic congestion. Algorithm 2 is adjusted to minimize the rate of production of travel related emissions. If a user-optimal assignment is one in which every user gets the best possible assignment they can get by any matching, a network optimal asks whether the total cost for all routes is minimized. For the network optimality result, we will begin with a simple reduction of the traffic assignment problem to online bipartite matching.

Consider a graph $G = (U, V, E)$, where the vertices in $U$ are the drivers arriving online and the vertices in $V$ are routes (Note that this is exactly online bipartite matching if we assume that there enough spots for all drivers that want to travel).

---

**ALGORITHM 2:** Online Driver-Route Matching

---

**for** *Each driver that comes online* **do**
    (1) Rank all routes available to driver d by cost $R(E, C)_r$
    (2) Assign top ranked route to driver $d$ with a deadline to begin travel
**end**

---

Karp et al. [8] shows that this ranking matching mechanism achieves a competitive ratio of $1 - \frac{1}{e}$.

## 4 COMPARISON TO AUCTION MECHANISM

This section compares our matching mechanism to an auction mechanism proposed byHeller et al. [7] for congestion pricing.Heller et al. [7] presents a two driver example to illustration their mechanism, we will show how the matching mechanism proposed here performs on the same scenario.

### 4.1 Auction mechanism

Under the auction mechanism (AUC), we have the same scenario of $n$ drivers traveling from Origin, $O$ to destination, $D$. For simplicity, but without loss of generality, we will assume a single route between $O$ and $D$ that may or may not be congested (as is assumed in Heller et al. [7], where the auction mechanism for congestion pricing is introduced). Every driver $d_i$ has a value to traveling at a certain time we call VOT ($\theta_i$) where ($\theta_i$) > 0. Every driver's VOT is unknown to the mechanism designer and drivers can travel(T) or opt out(O). Under the AUC, drivers can reduce travel time by paying higher prices which would translate to less drivers travelling. Under this mechanism, drivers place their bids according to their VOT and an allocation rule $A(x)$ determines which drivers get to travel, with those drivers paying a price determined by a payment rule $P(x)$.

Heller et al. [7] defines the utility of driver $i$ opting out to be

$$u_i^o = 0 \tag{2}$$

and traveling with VOT $\theta$, number of drivers $k$, and paying $p_i$, to be

$$u_i^T = v(\theta, k) - p_i, \tag{3}$$

where $v(\theta, k)$ is the value gained from the trip. This value can be gotten from $v(\theta, k) = \theta(S - c(k))$, where $S$ is time without traffic and $c(k)$ is the time with congestion.

### 4.2 The two driver example

Consider a scenario where we have two drivers that wish to travel with VOT ($\theta_1, \theta_2$)

From [7] the allocation rule would be as follows:

$$(x)\theta = \begin{cases} 1,1; if\ \theta_1 \in [\frac{1}{2}\theta_2, 2\theta_2] \\ 1,0; if\ [\theta_1 > 2\theta_2] \\ 0,1; if\ [\theta_1 < \frac{1}{2}\theta_2] \end{cases}$$

And payment rule:

$$p_1(\theta) = \begin{cases} 0; if\ [\theta_1 < \frac{1}{2}\theta_2] \\ \theta_2; if\ \theta_1 \in [\frac{1}{2}\theta_2, 2\theta_2] \\ 3\theta_2; if\ [\theta_1 > 2\theta_2] \end{cases}$$

then respective travel time times:

$$t_1(\theta) = \begin{cases} No\ travel; if\ [\theta_1 < \frac{1}{2}\theta_2] \\ 2; if\ \theta_1 \in [\frac{1}{2}\theta_2, 2\theta_2] \\ 1; if\ [\theta_1 > 2\theta_2] \end{cases}$$

From above we have three cases:

(1) case 1: $\theta_1 \in [\frac{1}{2}\theta_2, 2\theta_2]$
  (a) Allocations
  $$M_{AUC} = (1,1)$$
  $$M_{MAT} = (1,1)$$
  (b) Payments
  $$P_1^{AUC} = \theta_2$$
  $$P_1^{MAT} = C_d = \frac{i}{j}\theta_1 = \frac{i}{j}[\frac{1}{2}\theta_2, 2\theta_2]$$

(2) case 2: $[\theta_1 > 2\theta_2]$
  (a) Allocations
  $$M_{AUC} = (1,0)$$
  $$M_{MAT} = (1,1)$$
  (b) Payments
  $$P_1^{AUC} = 3\theta_2$$
  $$P_1^{MAT} = \frac{i}{j}\theta_1$$

(3) case 3: $[\theta_1 < \frac{1}{2}\theta_2]$
  (a) Allocations
  $$M_{AUC} = (0,1)$$
  $$M_{MAT} = (1,1)$$
  (b) Payments
  $$P_1^{AUC} = 0$$
  $$P_1 = \frac{i}{j}\theta_1$$

## 4.3  Utilities

Under some mild assumptions, we show here that the utilities achieved by the matching are superior to those achieved by the auction mechanism.

### 4.3.1  Utility under Auction.

$$U_d = \theta_d(t_f - k_c),$$

[7] assumes that,

$$t_c \equiv k_c, t_f = 4$$

therefore

$$U_d = \theta_d(4 - t_c)$$

### 4.3.2  Utility under Matching.

$$U_d = (t_f - t_c)C_d$$

but

$$C_d = \frac{i}{j}\theta_d$$

with

$$i < j, t_f = 4$$

then

$$C_d \le \theta_d$$

and therefore

$$U_d = (4 - t_c)\frac{i}{j}\theta_d$$

From above, we can see that the possible utility obtained from the auction mechanism by a driver, $d$ in the two driver example, is a fraction of the possible utility obtained under the matching mechanism, i.e;

$$U_d^{MAT} = \frac{i}{j}U_d^{AUC}$$

We can therefore conclude that the matching mechanism is preferable for each user since it achieves a higher payoff for the drivers,

$$U_d^{AUC} \le U_d^{MAT}$$

## 5 DISCUSSION

### 5.1 Implications

As Infrastructure-as-a Service, Platform-as-a Service and Software-as-a-Service become the dominant computing models for businesses large and small, Mobility-as-a-Service is emerging as an important, and in the future, possible dominant model for transportation. While the potential benefits of shared use autonomous vehicles are being examined from every available angle, the negative impacts of early MaaS systems (ride-hailing services in particular) are wreaking havoc in cities and at already congested sites including airports and rail terminals. Therefore, even in the US where there has been extensive resistance to congestion based pricing, cities such as New York, Los Angeles and San Francisco are considering following the leads of London, Singapore and Stockholm in enacting congestion pricing and low emissions zone pricing to reduce peak period congestion and pollution. Further, advances in tracking technology and emerging secure and privacy preserving contracting systems make it possible to tax vehicles or drivers for the use of limited infrastructure without putting their personal travel data at risk.

While estimates of the date at which society will see wide-scale adoption of autonomous vehicles have been widely inaccurate, someday this will surely be the norm. Research into the impact of autonomous vehicles on total vehicle-kilometer-traveled (VKT) have come up with highly variable estimates Fagnant and Kockelman [6], but what is assumed is that without road pricing that autonomous vehicles (many of the electric) would be free to impose externalities on the system paying only for fuel use. Therefore a matching mechanism such as ours would allocate scare resources more effectively.

### 5.2 Conclusion

This paper presents a matching mechanism for assigning drivers to routes where the drivers pay a toll for the marginal delay they impose on other drivers. The simple matching mechanism is derived from the RANKING algorithm for online bipartite matching proposed by Karp et al. [8]. The toll, which is anticipatory in design, is an adaption of one proposed by Dong et al. [5]. Our research proves that the matching mechanism proposed here is pareto user-optimal and can be adapted to give network optimal results for the minimizing total social cost of travel.

In the future, we intend to use a neural network model to predict short term future traffic flow conditions such as the one proposed by Asadi and Regan [2] and charge a toll that reflects those future conditions. Another goal is examine how efficient this matching mechanism is for solving the dynamic traffic assignment problem.